\documentclass[aps,prl,onecolumn,nofootinbib,groupedaddress]{revtex4-1}
\usepackage{amsmath,amsfonts,amssymb,bbm,epsfig}

\def\be{\begin{equation}}
\def\ee{\end{equation}}
\def\bq{\begin{equation}}
\def\eq{\end{equation}}
\def\bqa{\begin{eqnarray}}
\def\eqa{\end{eqnarray}}
\def\roughly#1{\mathrel{\raise.3ex
		\hbox{$#1$\kern-.75em\lower1ex\hbox{$\sim$}}}}
\def\lsim{\roughly<}
\def\gsim{\roughly>}
\def\llgm{\left\lgroup\matrix}
\def\rrgm{\right\rgroup}

\def\gslash#1{\slash\hspace*{-0.20cm}#1}
\renewcommand{\theequation}{\arabic{section}.\arabic{equation}}

\begin{document}
	\setlength{\oddsidemargin}{-0.03 cm}
	\def\bq{\begin{equation}}
	\def\eq{\end{equation}}
	\def\bqa{\begin{eqnarray}}
	\def\eqa{\end{eqnarray}}
	\def\roughly#1{\mathrel{\raise.3ex
			\hbox{$#1$\kern-.75em\lower1ex\hbox{$\sim$}}}}
	\def\lsim{\roughly<}
	\def\gsim{\roughly>}
	\def\llgm{\left\lgroup\matrix}
	\def\rrgm{\right\rgroup}
	\def\slash#1{\setbox\chek=\hbox{$#1$}\nude=\wd\chek#1{\kern-\nude/}}
	\def\gslash#1{\slash\hspace*{-0.20cm}#1}

	\begin{center}
		{\bf    Photoabsorption cross section in the low-$x$ and low-$Q^2$ domain, and DGLAP evolution }\\
		
	\end{center}
	\begin{center}
		G.R. Boroun \\

		Department of Physics, Razi University,
		Kermanshah 67149, Iran \\
		
		M. Kuroda\\
		Center for Liberal Arts, Meijigakuin University,
		Yokohama, Japan \\
		
		Dieter Schildknecht\\
		
		Fakult\"at f\"ur Physik, Universit\"at Bielefeld, 
		D-33501 Bielefeld, Germany\\
		
	\end{center}

\date{\today}
\vspace{0.5 cm}
\normalsize

{\bf Abstract}\\
The behavior of the gluon distribution of the proton in the low-$x$, low-$Q^2$ domain of deep inelastic electron-proton scattering (DIS) is being investigated.
By considering  two-gluon exchange as the dominant interaction in the low-$x$, low-$Q^2$ domain, we 
imply the well-known result of scaling of the photoabsorption cross section in terms of
the scaling variable $\eta(W^2,Q^2)$. From this,  we derive a reliable result for the 
gluon distribution at the leading order of the perturbative QCD improved parton model,  based on evolution from a starting scale of $Q_0^2\cong 2$  GeV$^2$.
The validity of evolution, when considering its quantitative modification at low-$Q^2$ 
without any alteration at larger values of $Q^2$, leads to a quantitative improvement in
the extraction of the gluon distribution based on evolution from a starting scale  of $Q^2$
 conventionally chosen as $Q^2= Q_0^2\cong 2$ GeV$^2$.

\setcounter{section}{1}
\section{1. Introduction}
The extraction of parton distribution functions  (PDF's) \cite{Harland, CJ, Ball, Buckley}
in the low-$x$, low-$Q^2$ domain\footnote{In conventional notation,$Q^2$ denotes 
the negative of the photon virtuality, and $x$ the Bjorken variable
$x\cong Q^2/W^2$, where $W^2$ refers to the photon-proton total energy squared.},
in particular, with the determination of the gluon distribution from measurements 
of deep inelastic scattering (DIS) in electron-proton scattering experiments \cite{Abramowicz} is interesting. The usually employed procedure rests on the choice of a "starting  scale", 
$Q_0^2$, at which 
a reasonable ansatz for the PDF's, in particular for the gluon density  as a function of $x$, containing a number of free parameters, is adopted  \cite{Harland, CJ, Ball}.  
Upon being evolved to a value of
$Q^2 \gtrsim Q_0^2$, and upon fit to the measured proton structure function at $Q^2>Q_0^2$, 
the above-mensioned free parameters in the ansatz at the starting scale  $Q_0^2$ 
are fixed. \footnote{The procedure allows one to include various different values 
of $Q^2$ within the single "global fit" in terms of the $x$-dependence 
at the $Q_0^2$ starting  scale. 
The error bars of the fit are considerably reduced compared to a fit 
restricted to a single value of $Q^2$.} 
The result on the input distribution at $Q^2=Q_0^2$, 
as emphasized by Pelicer at al., in ref.\cite{Pelicer}, obtained by different collaborations 
differs significantly  rather than being identical 
(within allowed small errors due to different fitting methods).
The fits are considered not to be satisfactory (compare to ref.\cite{Pelicer}). It seems worth noting that the resulting distribution at the input scale $Q_{0}^2$ 
does not obey any constraint in addition to
fulfilling the only and single purpose of determining the PDF's at values of 
$Q^2\gtrsim Q_0^2$; e.g.,  in particular, the $Q_0^2$ input distribution in general 
 differs significantly from the distribution required for a representation of 
the measured proton structure function at $Q^2\cong Q_0^2$.
The input distribution taken by itself is devoid  of any physical significance.

Our approach to the extraction of a reliable gluon distribution, to be elaborated upon
in the present paper,  is an entirely different one.  Our starting point is the
%the low-$x$, low-$Q^2$ 
(known) observation that DIS in the low-$x$, low-$Q^2$  domain is dominated by
two-gluon exchange to the (imaginary part of the) virtual Compton scattering amplitude.

The color-gauge-invariant ansatz for the two-gluon exchange amplitude leads to a
quantitative description of the experimental result for the (virtual) photoabsorption 
cross section $\sigma_{\gamma^*p}(W^2,Q^2)$ in terms of the low-$x$ 
scaling variable \cite{Schild},
\bq
   \eta(W^2,Q^2)=\frac{Q^2+m_0^2}{\Lambda^2_{sat}(W^2)},
\label{1.1}
\eq
i.e.
\bq   
     \sigma_{\gamma^*p}(W^2,Q^2) = \sigma_{\gamma^*p}(\eta(W^2,Q^2)).
\label{1.2}
\eq
 In (\ref{1.1}),      $\Lambda^2_{sat}(W^2)$  increases as a small power of $W^2$, and $m_0^2<m_\rho^2$, where $m_\rho$ denote the $\rho$-meson mass.  
 Compare   to Fig.2 \cite{Ku-Schi} and Section 2 for the $\eta(W^2,Q^2)$ dependence of the
 photoabsorption cross section.

The proton
structure function corresponding to  $\sigma_{\gamma^*p}(\eta(W^2,Q^2))$ will subsequently be used to derive the (dominant) gluon 
distribution. Compare to Section 3.
The accuracy and reliability of the extracted gluon distribution in the low-$x$, 
low-$Q^2$ domain at leading order of the pQCD improved parton model
will only be restricted by measurement errors and the associated reliability of the
photoabsorption cross section in terms of $\eta(W^2,Q^2)$.

The evolution of the proton structure function $F_2(x,Q^2)$ corresponding to 
$\sigma_{\gamma^*p}(\eta(W^2,Q^2))$  is examined in Section 4.
Standard evolution holds for  $\eta(W^2,Q^2)\gsim 1$, i.e. for fairly large values of 
$Q^2$, while requiring an explicitly quantitatively determined deviation for $\eta\lsim 1$.  
Compare to Section 4.

\setcounter{equation}{0}
\setcounter{section}{2}
\section{2. Basic assumptions of the color-dipole picture(CDP) and derivation of the
$\eta(W^2,Q^2)$ dependence}
It will be useful to work in the proton rest frame.  
The photon undergoes three-momentum conserving, energy-non-conserving transitions 
to on-shell $q\bar q$ states of finite mass,$M_{q\bar q}$\footnote{See e.g.Ref.\cite{KD} for a concise review of the well-known point}, and sufficiently 
long life time to interact with the proton. 

Consistency between the $Q^2$ dependence from $e^+e^-$ annihilation (entering the
$\gamma^*(q\bar q)$ transition) and $e^-p$ scattering  in the framework of the
underlying dispersion relation \cite{Sakurai} led to requiring opposite signs of interacting $q\bar q$ states neighboring in mass $M_{q\bar q}$  known as "off-diagonal generalized 
vector dominance"\cite{Fraas}.  In the framework of QCD, this opposite sign between 
contributions from neighboring states is recognized \cite{Cvetic} as a property of 
the color-gauge-invariant interaction of $q\bar q$ states with the proton 
via the structure of two-gluon couplings.  
Upon transition to the transverse distance between quark and antiquark, 
$\vec r_\perp$, one arrives at what is known as the "Color Dipole Picture (CDP)" \cite{Schild, Cvetic}.

\begin{figure}[h]
	\centering
	\begin{tabular}{cc}
		\includegraphics[width=8cm]{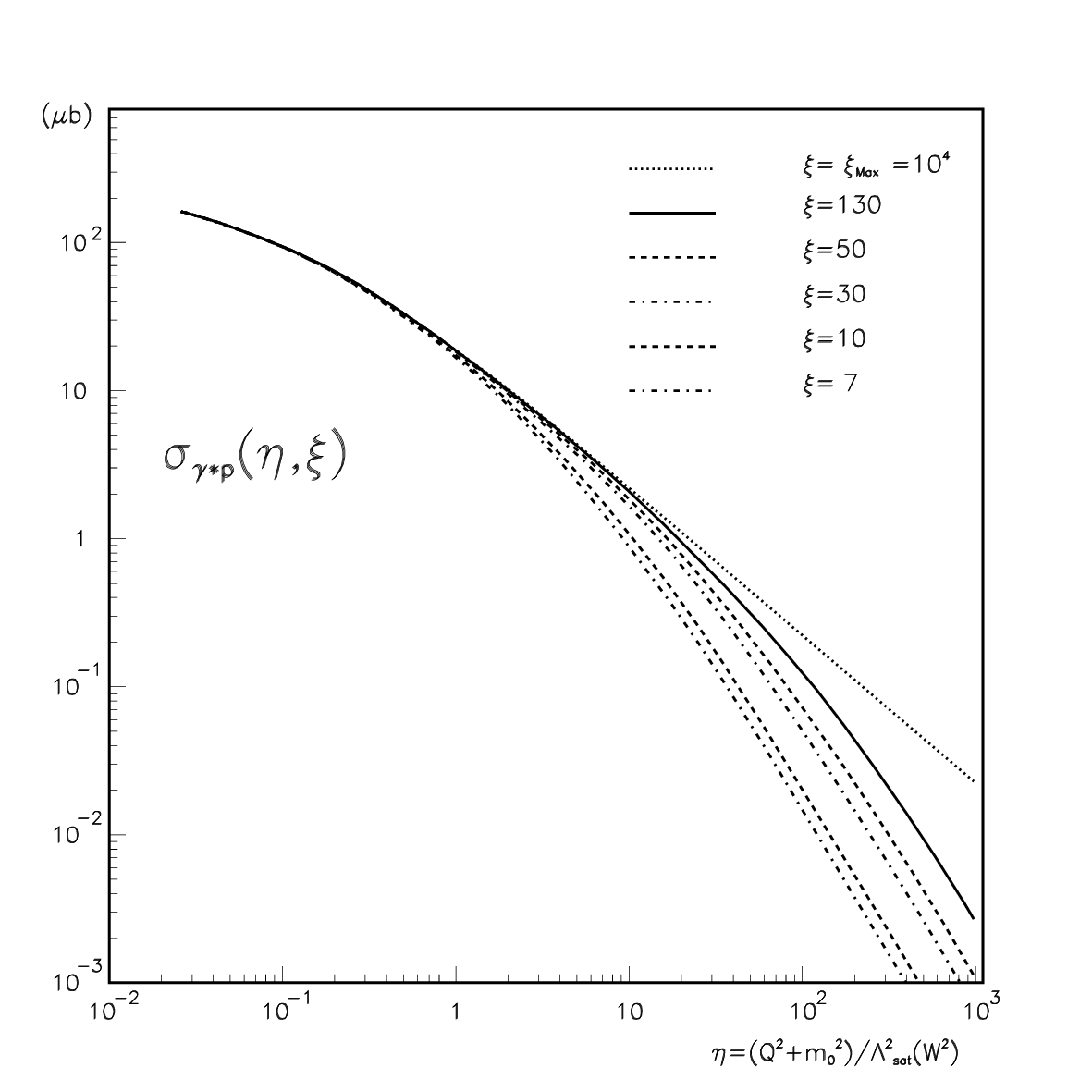}\\	
		\includegraphics[width=9cm]{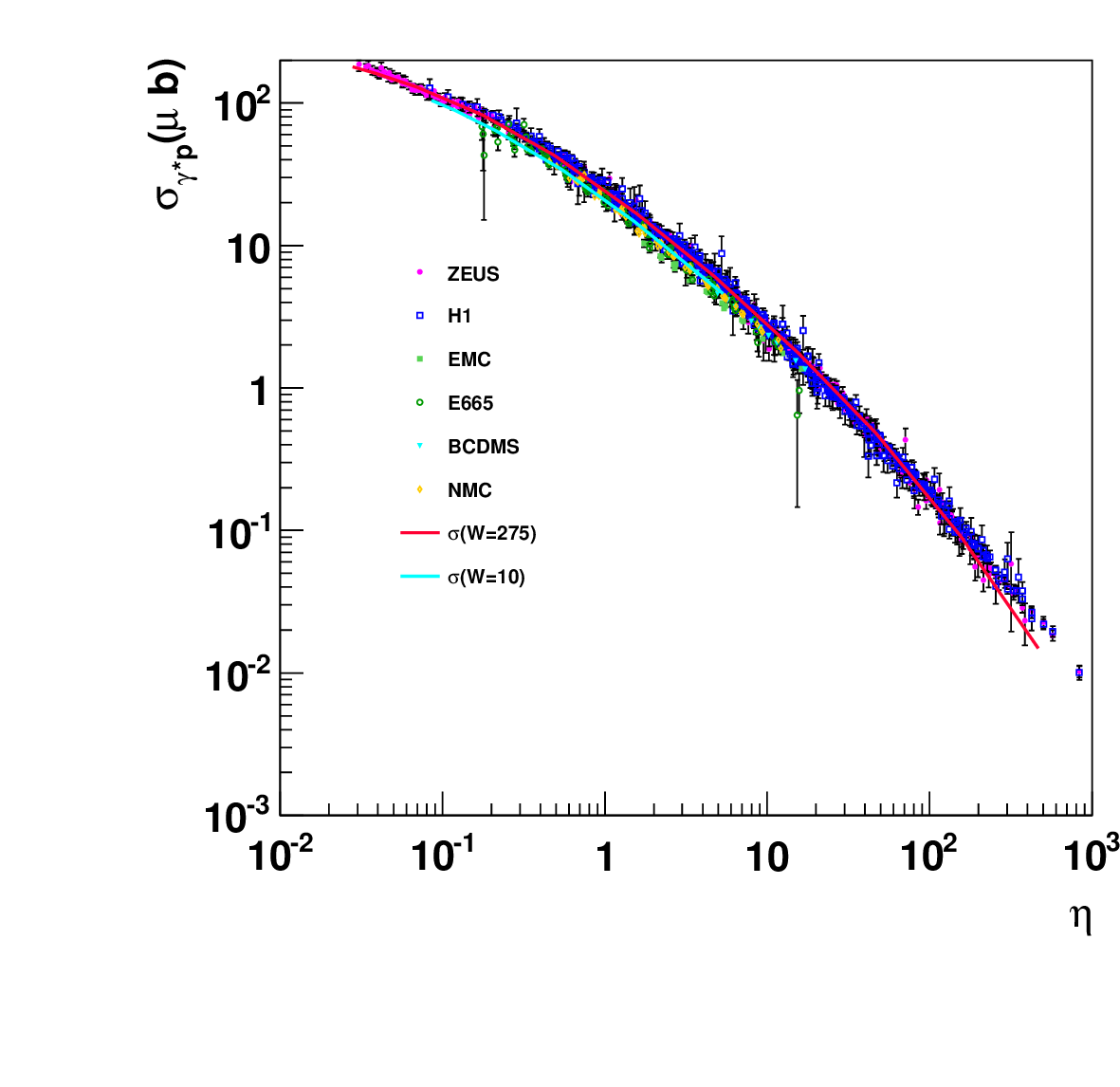}
	\end{tabular}
	\caption{\label{Fig1}The theoretical results for the photoabsorption cross section
		$\sigma_{\gamma^*p} (\eta (W^2,Q^2), \xi)$ in the CDP as a function of
		the low-x scaling variable $\eta (W^2,Q^2) = (Q^2 + m^2_0)/\Lambda^2_{sat}
		(W^2)$ for different values of the parameter $\xi$ that determines the
		(squared) mass range $M^2_{q \bar q} \le m^2_1 (W^2) = \xi \Lambda^2_{sat}
		(W^2)$ of the $\gamma^* \to q \bar q$ fluctuations that are taken into
		account. The experimental results   for 
		$\sigma_{\gamma^*p}(\eta(W^2,Q^2),\xi)$ (lower panel) lie on the full line corresponding
		to $\xi = \xi_0 = 130$, compare to ref. \cite{Ku-Schi}.
	}
\end{figure}

The basic assumption is the color-gauge invariant representation for the dipole 
interaction  resulting from two-gluon exchange \cite{Ku-Schi, Cvetic},
\bq
   \sigma_{(q\bar q)p}(\vec r_\perp, z(1-z),W^2)
   = \int d^2\vec\ell_\perp \tilde\sigma(\vec\ell_\perp^2, z(1-z),W^2)
       (1-e^{-i\vec\ell_\perp\vec r_\perp}),
\label{2.1}
\eq
entering 
\bq
       \sigma_{\gamma^*_{L,T}p} (W^2,Q^2)
  =   \int dz\int d^2\vec r_\perp\Big|  \Psi_{L,T}({\vec r_\perp}\hspace{0.01mm}^2, z(1-z),Q^2)\Big|^2
        \sigma_{(q\bar q)p}(\vec r_\perp, z(1-z),W^2).
\label{2.2}
\eq
We note tht the color-gauge-invariant representation (\ref{2.1}) by itself is sufficient \cite{Ku-Schi} to derive the low-$\eta(W^2,Q^2)$ scaling behavior,
\bq
     \sigma_{\gamma^*p}(W^2,Q^2)=\sigma_{\gamma^*p}(\eta(W^2,Q^2)),
\label{2.3}
\eq
with $\Lambda_{sat}^2(W^2)$ being given by
\bq
       \Lambda^2_{sat}(W^2)=\frac{\int d\vec\ell_\perp^{\prime 2} \vec\ell_\perp^{\prime 2} 
          \tilde\sigma_{(q\bar q)_L^{J=1}p} (\vec\ell_\perp^{\prime 2},W^2)}
          { \int d\vec\ell_\perp^{\prime 2}
          \tilde\sigma_{(q\bar q)_L^{J=1}p} (\vec\ell_\perp^{\prime 2},W^2)},
\label{2.4}
\eq
$$
     \vec\ell_\perp^{\prime \hspace{0.5mm}2} \equiv {{\vec\ell_\perp}\hspace{0.01mm}^{2}\over{z(1-z)}},
$$     
as well as the low-$\eta$ and the large-$\eta$ dependence of the photoabsorption 
cross section,
\bq
       \sigma_{\gamma^* p} (W^2, Q^2)=\sigma_{\gamma^* p} (\eta(W^2, Q^2))
    =   {\alpha\over \pi}\sum_qQ_q^2 \left\{\begin{array} {l@{\quad,\quad}l} 
               \sigma_T^{(\infty)}(W^2)\ln{1\over{\eta(W^2,Q^2)}} &  \eta\ll 1, \\
               {1\over 6}(1+2\rho)\sigma_L^{(\infty)}(W^2) {1\over\eta} & 1 \ll\eta\lsim 40. \end{array} \right. 
 \label{2.5}
 \eq
In (\ref{2.5}), 
\bq
   \sigma_T^{(\infty)}(W^2)=\pi\int d\vec\ell_\perp^{\prime 2} \tilde\sigma_{(q\bar q)_T^{J=1}p}
       (\vec\ell_\perp^{\prime 2},W^2)
       =\rho\sigma_L^{(\infty)}(W^2)=\rho \pi   \int d\vec\ell_\perp^{\prime 2} 
       \tilde\sigma_{(q\bar q)_L^{J=1}p}(\vec\ell_\perp^{\prime 2},W^2),           
\label{2.6}
\eq
where $\sigma_T^{(\infty)}(W^2)$ and $\sigma_L^{(\infty)}(W^2)$ are weakly $W^2$-dependent cross
section for $(q\bar q)_{T,L}$-proton scattering of hadronic size.  A preferable value of
$\rho$ in (\ref{2.6}) is given by $\rho={4\over 3}$. The weakly $W^2$ dependent cross section (\ref{2.6}) is  related to and is replaced by reliable independent fits to
$Q^2=0$ photoproduction.  The factor $\rho$ in (\ref{2.6}), via $R(W^2,Q^2)=1/2\rho$ 
determines the longitudinal-to transverse ratio, $R(W^2,Q^2)$ of the photoabsorption 
cross section $\sigma_{\gamma^* p} (\eta(W^2, Q^2))$.

Upon specification of $\tilde\sigma(\vec\ell_\perp^2,z(1-z),W^2)$ in  (\ref{2.1}) via
\bq
\tilde \sigma (\vec l^{~2}_\bot, z (1-z), W^2) = 
\frac{\sigma^{(\infty)} (W^2)}{\pi}
\delta \left(\vec l^{~2}_\bot - z (1-z) \Lambda^2_{sat} (W^2) \right),
\label{2.7}
\eq
one obtains the full CDP results in (2.3) shown in Fig.1.  Compare to Appendix A for 
the derivation  of the analytic expression for $\sigma_{\gamma^* p} (\eta(W^2, Q^2))$.

The parameters entering the result in Fig.1 are given by
\bq
    m_0^2=0.15 {\rm GeV}^2
\label{2.8}
\eq
as well as
\bqa
        C_1&=& 0.31 {\rm GeV}^2,   \nonumber \\
        C_2 &=& 0.27,
 \label{2.9}
 \eqa
 where\footnote{Validity of evolution at large $Q^2$ implies the constraint $C_2= 0.29$ consistent with experiment. See Appendix B for this known restriction on the exponent $C_2$ 
 	that reduces the fitted parameters to $m_0^2$ and the normalization $C_1$. }
 \bq
    \Lambda^2_{sat}(W^2)=C_1\Bigl({{W^2}\over{1{\rm GeV}^2}}\Bigr)^{C_2}.
 \label{2.10}
 \eq
The results may equivalently be expressed in terms of the proton
structure function
\bq
   F_2 (W^2, Q^2) =
       \frac{Q^2}{4 \pi^2 \alpha} \left( \sigma_{\gamma^*_T p} (W^2, Q^2) +
\sigma_{\gamma^*_L p} (W^2, Q^2)\right)
     =  \frac{Q^2}{4 \pi^2 \alpha}  \sigma_{\gamma^* p} (\eta(W^2, Q^2)). 
\label{2.11}
\eq 

We note the range of $\eta(W^2,Q^2)$ for the value of $Q^2=1.9$GeV$^2$ and
presently relevant values of $W^2=10^3$ GeV$^2$ to $W^2=10^5$GeV$^2$ that
is given by
\bq
    0.30\lsim \eta(W^2,Q^2) \lsim 1.03.
\label{2.12}
\eq
The parameter $\xi$ also used in Fig.1 yields the mass range of $q\bar q$ transitions,
\bq
     M_{q\bar q}^2\le m_1^2 =\xi \Lambda_{sat}^2(W^2).
\label{2.13}
\eq           
The finite value of $\xi$ only becomes relevant for $\eta(W^2,Q^2) \gsim 10$.
For $\eta(W^2,Q^2) \lsim 10$, $\xi$ may be put to $\xi=\infty$.
In the transition to small $\eta(W^2,Q^2)$, i.e. small $Q^2$,  the transition of the photon to
large values of $M_{q\bar q}$ dies out.  Compare to Fig.1.

\setcounter{equation}{0}
\setcounter{section}{3}
\section{3. The gluon distribution function}
In leading order of pQCD, the longitudinal structure function, $F_L(x,Q^2)$, 
as a consequence of the dominance of the gluon distribution at low-$x$, is determined by \cite{Martin}
\bq
      F_L (x,Q^2)= \frac{2 \alpha_s (Q^2)}{\pi} \left(\sum_q Q^2_q\right) I_g(x,Q^2),
\label{3.1}
\eq
where $I_g(x,Q^2)$ denotes an integral over the gluon distribution,
\bq
    I_g(x,Q^2) = \int^1_x \frac{dy}{y} \left( \frac{x}{y} \right)^2 \left(1 - \frac{x}{y} \right) G (y, Q^2).
\label{3.2}
\eq
For a considerable range of gluon distributions, $G(y,Q^2)$, the integral in {\ref{3.2} 
may be replaced by the simple proportionality
 \bq
   F_L \left(\xi_L x, Q^2\right) = \frac{\alpha_s(Q^2)}{3 \pi}\left( \sum_q Q^2_q\right) G (x,Q^2),
\label{3.3}
\eq
where the shift parameter $x\to \xi_L x$ on the left-hand side in (\ref{3.3}) has 
 the preferred value of $\xi_L\cong 0.40$ \cite{Martin}.

The CDP result for the longitudinal structure function is compared to available 
experimental results in Fig.2.

 \begin{figure}[h]
 	\includegraphics[width=12cm]{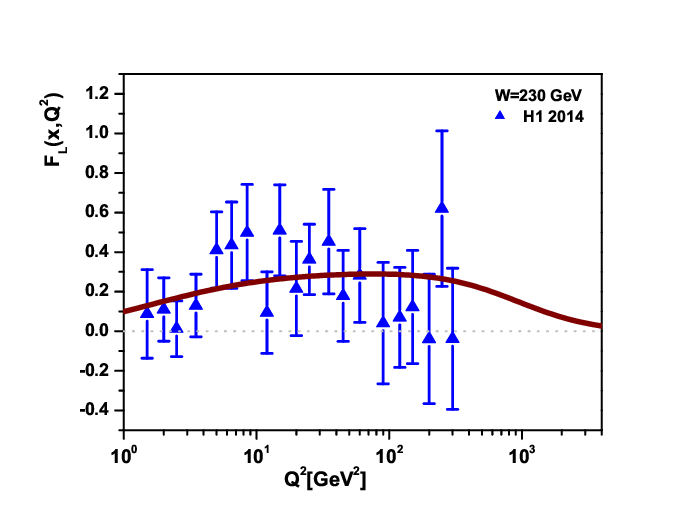}%
 	\caption{\label{Fig2} Experimental data \cite{Andreev} and the CDP prediction of 
	$F_L(x,Q^2)$ are	compared.}
 \end{figure}

Substitution of the CDP result for $F_L(\xi_Lx,Q^2)$ into (\ref{3.3}) allows one to determine the gluon distribution,
$G(x,Q^2)$, in terms of $F_L(\xi_L x,Q^2)$.
The accuracy of the gluon distribution resulting from $F_L(\xi_L x,Q^2)$ is tested 
by insertion of $G(x,Q^2)$ obtained from (\ref{3.3})   into the integral in (\ref{3.1}).

The resulting $F_L(x,Q^2)$ agrees with the input for $F_L(x, Q^2)$ within less than 4\%, compare to Table 1. The gluon distribution obtained from (\ref{3.3}) for $1~\text{GeV}^2{\lesssim}Q^2{\lesssim}100~\text{GeV}^2$ and $10^3~\text{GeV}^2{\lesssim}W^2{\lesssim}10^5~\text{GeV}^2$ indeed reproduces $F_L(x,Q^2)$ in (\ref{3.1}).

 \begin{table}[h]
 	\begin{tabular}{|l|c|c|c|}
 		\hline
 		& $W^2 = 10^5$ & $W^2 = 10^4$ & $W^2 = 10^3$ \\
 		& $[ {\rm GeV}^2 ] $ & $[ {\rm GeV}^2 ] $ & $[ {\rm GeV}^2 ] $ \\
 		\hline
 		$Q^2 = 100$ & 1.043 & 1.037 & 0.981 \\
 		$[ {\rm GeV}^2 ]$ & & & \\
 		\hline
 		$Q^2 = 50$ & 1.041 & 1.037 & 1.012 \\
 		$[ {\rm GeV}^2 ]$ & & & \\
 		\hline
 		$Q^2 = 10$ & 1.035 & 1.035 & 1.030 \\
 		$[ {\rm GeV}^2 ]$ & & & \\
 		\hline
 		$Q^2 = 2$ & 1.021 & 1.026 & 1.029 \\
 		$[ {\rm GeV}^2 ]$ & & & \\
 		\hline
 		$Q^2 = 1$ & 1.015 & 1.021 & 1.026 \\
 		$[ {\rm GeV}^2 ]$ & & & \\
 		\hline
 	\end{tabular}
 	\caption{The table shows the ratio of $F_L (W^2,Q^2)$ in (\ref{3.1}) to $F_L(W^2,Q^2)$
 		from the CDP, where the former is calculated
 		by inserting $G(x,Q^2)$  obtained from (\ref{3.3}) by using $F_L(W^2,Q^2)$
 		from the CDP.}
 \end{table}

The derived gluon distribution is relevant for $Q^2\gsim Q_0^2$, where $Q_0^2$ must 
be large enough to allow for $\alpha_s(Q^2) \ll 1$. Our method of determining the gluon distribution at leading order of pQCD from the successful fits to 
the experimental results for the proton structure functions in the CDP yields 
unambiguous and reliable results
for the gluon distribution of the proton  at low-$x$ and chosen values of $Q^2$. Indeed, evaluation of the pQCD result for $F_L(x,Q^2)$ in (3.1) in terms of the extracted gluon distribution reproduces the experimentally determined longitudinal structure function of the proton, see Table I.

%In Fig.4, for the very specific choice of $Q^2=Q_0^2\cong 1.9$GeV$^2$, the resulting gluon 
%distribution is compared to the large spread of distributions from the big collaborations
%taken from Fig.1.
%The error of our result at leading order of pQCD for the gluon at low $x$ for given $Q^2$ is %essentially 
%restricted to the small error of the CDP theoretical result in the $\eta(W^2,Q^2)$ plot of Fig.2 %compared to the shown experimental results.

%\begin{figure}[h]
%	\includegraphics[width=12cm]{Fig4.eps}%
%	\caption{\label{Fig4} The CDP results for $G(x,Q^2)=xg(x,Q^2)$ are compared with the results %from ref.[1] and the
%		parametrization methods,  CJ12 \cite{CJ}, NNPDF3.0 \cite{Ball} and
%		MMHT14 \cite{Harland} as accompanied with total errors.
	%}
%\end{figure}

\setcounter{section}{4}
\setcounter{equation}{0}
\section{4. Evolution}
We turn to examining evolution, i.e. studying the logarithmic derivative
$\frac{\partial F_2(x,Q^2)}{\partial \ln Q^2}$ of the experimentally determined structure function $F_2(x,Q^2)$.

In the low-$x$, low-$Q^2$ domain of the pQCD improved parton model for $Q^2>>1~\text{GeV}^2$, at leading order,
DIS is dominated by the gluon distribution $G(x,Q^2)$.  Both $F_L(x,Q^2)$ as well as 
$\frac{\partial F_2(x,Q^2)}{\partial \ln Q^2}$ are related to $G(x,Q^2)$ by integrals of similar magnitude \cite{Martin},
\bqa
    F_L(x,Q^2) &=& {{2\alpha_s(Q^2)}\over \pi}\int_x^1{{dy}\over y} \Bigl({x\over y}\Bigr)^2
         \Bigl( 1-{x\over y}\Bigr) G(y,Q^2)  \nonumber \\
       &\cong& {{10\alpha_s(Q^2)}\over{27 \pi}} G({x\over{\xi_L}},Q^2)
\label{4.1}
\eqa
with
\bq
      {1\over{\xi_L}}={1\over{0.4}}=2.5,
\label{4.2}
\eq
and \cite{Prytz}                 
\bqa
   \frac{\partial F_2 (x,Q^2)}{\partial \ln Q^2} &=& {{5\alpha_s}\over{9\pi}}\int_0^{1-x}dz
         \Bigl( (1-z)^2+z^2 \Bigr) G({x\over{1-z}},Q^2)   \nonumber \\
         &\cong&  {{10\alpha_s(Q^2)}\over{27\pi} }G({x\over{\xi_2}}),
\label{4.3}
\eqa
where
\bq
      {1\over{\xi_2}}={1\over{0.5}}=2.0.
\label{4.4}
\eq
 According to (\ref{4.3}) and (\ref{4.1}),
 \bqa
     \frac{\partial F_2 (x,Q^2)}{\partial \ln Q^2}  &=& 
          F_L({{\xi_L}\over{\xi_2}}x,Q^2)  \nonumber \\
         &\cong& F_L(x,Q^2),
 \label{4.5}
 \eqa
where the shift factor ${{\xi_L}\over{\xi_2}}=0.8$
was approximated by  ${{\xi_L}\over{\xi_2}}=1$ in the second step of (\ref{4.5}). In the pQCD improved parton model, in terms of the underlying gluon distribution, evolution (\ref{4.5}) is recognised  as a consequence of the proportionalities (\ref{4.1}) and (\ref{4.3}).

In  distinction from the pQCD improved parton model, where the range of $Q^2$ in (\ref{4.1}) and (\ref{4.3}) must be 
restricted to sufficiently large values of $Q^2 \gg 1$ GeV$^2$, in the CDP 
the representation of the measured photoabsorption cross section 
$\sigma_{\gamma^*p}(\eta(W^2,Q^2))$ and the associated structure function 
include the transition to low $Q^2$, including the $Q^2=0$ photoproduction limit.
The CDP allows one to quantitatively analyze the magnitude of the potential deviation 
from DGLAP evolution, when passing from the validity of standard evolution  at large 
$Q^2$ to the low-$Q^2$ domain.

Differentiation of $F_2(x,Q^2)$ in (\ref{2.11}), upon replacing $W^2$ by $W^2={{Q^2}\over x}$,
and simplifying the notation via
\bq
     \sigma_{\gamma^*_T p}(x,Q^2) =\sigma_T,~~~~ 
      \sigma_{\gamma^*_L p}(x,Q^2) =\sigma_L,
\label{4.6}
\eq
leads to 
\bqa
   \frac{\partial F_2 (x,Q^2)}{\partial \ln Q^2} &=& F_L(x,Q^2) \Bigl( 1+{{\sigma_T}\over{\sigma_L}}
   +{{Q^2}\over{\sigma_L}}{\partial\over{\partial Q^2}}(\sigma_T
   +\sigma_L)\Bigr)    \nonumber \\
   &\equiv & F_L(x,Q^2) R_3(x,Q^2),
\label{4.7}
\eqa
Comparison of (\ref{4.7}) and the second line of (\ref{4.5})
shows that $R_3(x,Q^2)$ indeed quantifies deviations from DGLAP evolution
($R_3(x,Q^2)=1$), when passing from the large-$Q^2$ validity to
the low-$Q^2$ domain.

Equation (\ref{4.7}) is evaluated by substitution of the CDP results given in Appendix A, i.e.
\bq
   F_L(x,Q^2)={{Q^2}\over{4\pi^2\alpha}}
          \frac{\sigma_{\gamma p}(W^2)} { \Bigl(\ln {\rho\over{\mu(W^2)}}\Bigr) }
          I_L^{(1)}(\eta,\mu),
\label{4.8}
\eq
and $R_3(x,Q^2)$ is given by
 \bq
   R_3(x,Q^2)=\frac {1}{I_L^{(1)}\Bigl(\eta,\mu\Bigr)}
      \Bigl[ I_T^{(1)}({\eta\over\rho},{\mu\over\rho}) +I_L^{(1)}(\eta,\rho)
         +{\partial\over{\partial\ln Q^2}} 
           \Bigl(  I_T^{(1)}({\eta\over\rho} {\mu\over\rho}) +I_L^{(1)}(\eta,\rho)\Bigr) \Bigr],
\label{4.9}
\eq
where
\bqa
    I_L^{(1)}(\eta,\mu)&=&  {{\eta-\mu}\over\eta}\Bigl(1-2\eta I_0(\eta) \Bigr),  \nonumber \\
    I_T^{(1)}(\eta,\mu) &=&  I_0(\eta)- {{\eta-\mu}\over\eta} \Bigl(1-2\eta I_0(\eta) \Bigr), 
\label{4.10}
\eqa
with
\bq
    I_0(\eta) = {1\over{\sqrt{1+4\eta}}}\ln\frac{\sqrt{1+4\eta}+1}{\sqrt{1+4\eta}-1}.
\label{4.11}
\eq
The results for $R_3(x,Q^2)$ as a function of $Q^2$ for various choices of $x$ are 
presented in Fig.3.  
At large $Q^2$, $R_3(x,Q^2)$ converges towards unity, $R_3(x,Q^2)=1$, corresponding to
validity of standard evolution according to (\ref{4.5}) and (\ref{4.7}). 

\begin{figure}[h]
	\includegraphics[width=12cm]{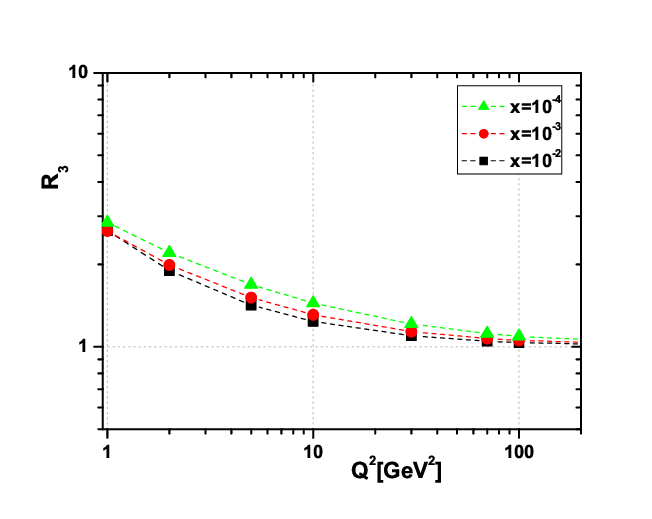}%
         \caption{\label{Fig3}  The factor $R_3(x,Q^2)$ in (\ref{4.7}), multiplying $F_L(x,Q^2)$ in the CDP, as a function of $Q^2$ 
         at fixed values of $x$.}
\end{figure}

For low values of $Q^2$, in Fig.3, one observes an increasingly stronger rise of $R_3(x,Q^2)$ corresponding to a correction factor to standard evolution that is given by $R_3(x,Q^2)$.

The increase of $R_3(x,Q^2)$ with deceasing $Q^2$ may be directly visualized as a multiplicative factor to $\sigma_{\gamma^*p}(\eta(W^2,Q^2))$.  Indeed \cite{BKS},
 \bq
    \frac{\partial F_2 (x,Q^2)}{\partial \ln Q^2} ={{Q^2}\over{4\pi^2\alpha}}{1\over{1+2\rho}}
      \sigma_{\gamma^*p}(\eta(W^2,Q^2)_{ W^2=Q^2/ x})R_3(x,Q^2).
\label{4.12}
\eq

or

 \bq
     \sigma_{\gamma^*p}(\eta(W^2,Q^2)_{ W^2=Q^2/ x})=\frac{\partial F_2 (x,Q^2)}{\partial \ln Q^2} {{4\pi^2\alpha}\over{Q^2}}{(1+2\rho)}
  {1\over{ R_3(x,Q^2)}}.
\label{4.13}
\eq
In Fig.4, we show the result for $\sigma_{\gamma^*p}(\eta(W^2,Q^2))$ from 
Fig.1, along with the impact of  multiplying by $R_3(x,Q^2)$. It is evident that a factor of $1/R_3(x,Q^2)$ in Eq. (\ref{4.13}), where $R_3(x,Q^2)$ is defined in Eq. (\ref{4.9}), which is different from unity,  must be applied to the evolution prediction at low $Q^2$ ($\eta{\lesssim}1$) to achieve consistency between evolution and the experimental results. The experimental data, represented by the $\eta(W^2, Q^2)$-dependence in Fig.1 from the CDP, satisfy evolution ($R_3(x,Q^2)=1$). The discrepancy between the experimental data (black-solid curve) and the DGLAP evolution (red-dashed curve) at low $\eta$ indicates that the relatively low initial scale  used in many global analyses to extract proton PDFs is too low, and the DGLAP evolution cannot be relied upon in that region. 
By using $\eta$, we observe a large $\eta$ behavior as $1/\eta$ and a deviation from $1/\eta$ for $\eta$ less than 1. Since DGLAP implies $1/\eta$, this principle is violated at a scale of approximately $\simeq{1.9~\text{GeV}^2}$.  Therefore, $Q^2\simeq{1.9~\text{GeV}^2}$ can not be used as a starting scale. The data exhibit inconsistency with evolution at low $\eta$.

\begin{figure}[h]
	\includegraphics[width=10cm]{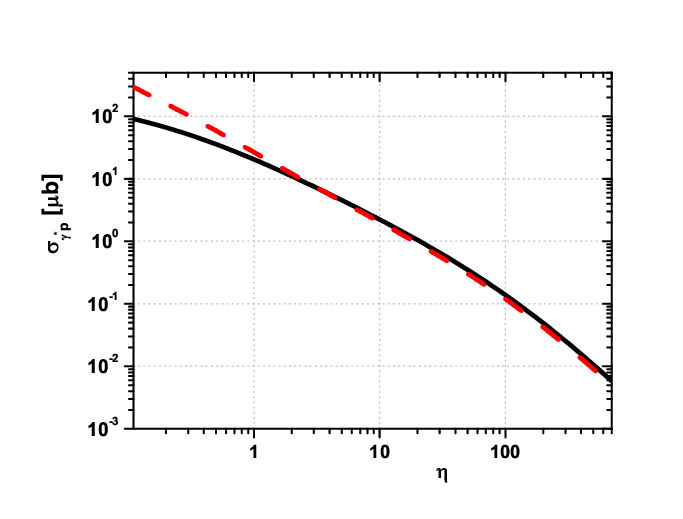}%
	\vspace {1.5 cm}
         \caption{\label{Fig4} The photoabsorption cross section (\ref{1.2})
         $\sigma_{\gamma^*p}(\eta ={{Q^2+m_0^2}\over{\Lambda_{sat}^2}})$ (black curve) describing the DIS experimental results  
          compared to $\sigma_{\gamma^*p}(\eta)R_3(x,Q^2)$ according to (\ref{4.12}) 
          (red curve).}
\end{figure}

\setcounter{section}{5}
\setcounter{equation}{0}
\section{5. Conclusion}
Based on the dominant interaction of two-gluon exchange  in photoabsorption and 
the associated structure functions of the proton in DIS, which implies scaling in terms of 
the low $x$ scaling  variable $\eta(W^2,Q^2)$, we have derived a reliable  
result for the gluon distribution function at the leading order of pQCD in the low-$x$, low-$Q^2$ domain.
The DGLAP evolution is consistent with experimental results for $F_2(x,Q^2)$ in the CDP at large $Q^2$.  We provide explicit quantitative 
results for its necessary modification at low $Q^2$, including
the behavior as $Q^2$ approaches  very low values towards $Q^2=0$.  
With this modification, the evolution is applicable to experimental data of
DIS in the CDP at any $Q^2$ in the low-$x$, low-$Q^2$ domain.

Our results suggest a method for determining of the gluon distribution by evolving from a given starting scale $Q^2$ approximately $Q_0^2\cong 2$ GeV$^2$.
The assumed value at the starting scale should be supplemented by the experimentally determined DIS result 
for the structure function at the value of $Q_0^2$.  
Evolution of the structure function from $Q_0^2$ to
$Q^2\gsim Q_0^2$, using the mentioned low $Q^2$ modification, will lead to 
a successful result for the structure function at large $Q^2$ and a reliable result 
for the associated gluon distribution.

\renewcommand{\theequation}{\Alph{section}.\arabic{equation}}
\setcounter{section}{1}
\setcounter{equation}{0}
\section{Appendix A.  Analytic expression for $\sigma_{\gamma^*p} (W^2,Q^2)$ and for $F_L(x,Q^2)$ in the CDP.}
The  transverse and longitudinal photoabsorption cross section in CDP is given by \cite{{Ku-Schi}}
\bqa
\sigma_{\gamma^*_Lp} (W^2,Q^2) &=& \frac{\alpha R_{e^+e^-}}{3 \pi}
\sigma^{(\infty)} (W^2) I_L \left( \eta (W^2,Q^2), \mu (W^2) \right) G_L(u).
\label{a.1}\\
\sigma_{\gamma^*_Tp} (W^2,Q^2) &=& \frac{\alpha R_{e^+e^-}}{3 \pi}
\sigma^{(\infty)} (W^2) I_T \left( {{\eta (W^2,Q^2)}\over\rho}, {{\mu (W^2)}\over\rho} \right) G_T(u),
\label{a.2}
\eqa

where
\bqa
\mu(W^2)=\eta(W^2,0) =\frac{m_0^2}{\Lambda_{sat}^2(W^2)}.
\label{a.3}
\eqa

In (\ref{a.1}) and (\ref{a.2}), $R_{e^+e^-} = 3 \sum_q Q^2_q$, where the sum runs over
the actively contributing quark flavors, and $Q_q$ denotes the quark charge, 
while  $G_T(u)$ and $G_L(u)$ are correction factors due to the finite value of $\xi$ with
\bq
u={\xi\over \eta}.
\label{a.4}
\eq
The quantity $\rho ={4\over 3}$ is the ratio of the size of the $(q\bar q)$ dipole  in 
the transeverse and ongitudinal amplitude,
which signifies the enhancement of dipole size in the transverse amplitude. 
It turns out that the relevant region of $\mu (W^2)$ fulfills the bound
$\mu (W^2) < 1$. Under this assumption $I_{L,T} (\eta (W^2,Q^2), \mu
(W^2))$ becomes 
\be
I_{L,T} (\eta (W^2,Q^2),\mu (W^2)) = I^{(1)}_{L,T} (\eta (W^2,Q^2), \mu (W^2))
\left( 1+0 \left( \mu (W^2)\right) \right),
\label{a.5}
\ee
where $I^{(1)}_L (\eta (W^2,Q^2), \mu (W^2))$ and $I^{(1)}_T 
(\eta (W^2,Q^2), \mu (W^2))$
are given by
\bqa
&& I^{(1)}_L (\eta, \mu) = \frac{\eta - \mu}{\eta} \nonumber \\
&& \times \left( 1 - 
\frac{\eta}{\sqrt{1+4 (\eta - \mu)}} \ln 
\frac{\eta (1+ \sqrt{1+4(\eta - \mu)})}{4 \mu-1-3\eta + 
	\sqrt{(1+4(\eta - \mu))((1+\eta)^2 - 4 \mu)}} \right), 
\label{a.6} \\
&& I^{(1)}_T (\eta, \mu)  =  \frac{1}{2} \ln \frac{\eta-1 + 
	\sqrt{(1+\eta)^2 - 4 \mu}}{2 \eta} 
- \frac{\eta - \mu}{\eta} + \frac{1+2(\eta - \mu)}
{2 \sqrt{1+4 (\eta - \mu)}}  \nonumber \\
&&~~~~~~~~~~~~~ \times  \ln \frac{\eta (1 + \sqrt{1+4 (\eta - \mu)})}
{4 \mu -1-3\eta + \sqrt{(1+4(\eta-\mu))((1+\eta)^2-4 \mu)}}.
\label{a.7}
\eqa

We note the photoproduction $(Q^2 = 0)$ limit of (\ref{a.2}),
%
% with
%(\ref{2.9}). Inserting $\eta = c \mu (W^2)$ into (\ref{2.9}), where
%$c = const \ge 1$ and $0 < \mu (W^2) < 1$, a careful evaluation of the
%photoproduction limit of $c \to 1$  
\bq
\sigma_{\gamma p} (W^2) \equiv  \lim_{\eta \to \mu} \sigma_{\gamma^*_Tp}(W^2,Q^2) 
= \frac{\alpha R_{e^+e^-}}{3 \pi} \sigma^{(\infty)} (W^2) G_T(u)\ln\frac{\rho}{\mu}.
\label{a.8}
\eq

In the limit of very high energy, $\mu (W^2) \ll 1$, (\ref{a.5}) 
may be further simplified.
We note that $\mu (W^2) \le \eta (W^2,Q^2) \le \eta_{Max} (W^2)$, 
where $\eta_{Max} (W^2)$ is
determined by the required restriction to low values 
of $x \le x_0 \lsim 0.1$, or
$Q^2 \le x_0W^2$, i.e.
\be
\eta(W^2,Q^2) \le \eta_{Max} (W^2) = \frac{x_0W^2}{\Lambda^2_{sat} (W^2)}.
\label{a.9}
\ee
With $\mu (W^2) \ll 1$ and $\eta (W^2,Q^2) \ge \mu (W^2)$, 
upon making use of the identity
\bqa
2 \ln \frac{\sqrt{1+4 \eta} +1}{\sqrt{1+4 \eta} -1} & = & \ln
\frac{(1+\eta) \sqrt{1+4 \eta} + 1 + 3 \eta}{\eta (\sqrt{1 + 4 \eta} -1)} =
\nonumber \\
& = & \ln \frac{\eta (1 + \sqrt{1 + 4 \eta})}{(1 + \eta) \sqrt{1 + 4\eta}
	-1 - 3 \eta},
\label{a.10}
\eqa
and of the definition
\be
I_0 (\eta) = \frac{1}{\sqrt{1 + 4 \eta}} \ln 
\frac{\sqrt{1 + 4 \eta} +1}{\sqrt{1 + 4 \eta} -1},
\label{a.11}
\ee
we find that (\ref{a.5}) and (\ref{a.6})  become
\bqa
I^{(1)}_L (\eta, \mu) & = & \frac{\eta - \mu}{\eta} (1 - 2 \eta I_0 (\eta)),
\nonumber \\
I^{(1)}_T (\eta, \mu) & = & I_0 (\eta) - \frac{\eta - \mu}{\eta}
(1 - 2 \eta I_0(\eta)).
\label{a.12}
\eqa
$\sigma^{(\infty)}(W^2)$ in (\ref{a.1}) and (\ref{a.2}) is evaluated from the photoproduction 
limit  (\ref{a.8}) as
\bq
\sigma^{(\infty)}(W^2) = {{3\pi}\over{\alpha R_{e^+e^-}}} 
{1\over{\Bigl(\ln {\rho\over \mu}\Bigr) G_T(u)}} 
\sigma_{\gamma p}(W^2).
\label{a.13}
\eq
Inserting (\ref{a.13}) in (\ref{a.1}), we obtain

\bq
F_L(x,Q^2)=  {{Q^2}\over{4\pi^2\alpha}} \sigma_{\gamma^*_L p}(W^2, Q^2)
= {{Q^2 }\over{4\pi^2\alpha}}
\frac{\sigma_{\gamma p}(W^2)} { \Bigl(\ln {\rho\over{\mu(W^2)}}\Bigr) G_T(u)}
I_L^{(1)}(\eta,\mu)G_L(u),
\label{a.14}
\eq
Since $\xi\cong 130$, while $\eta \lsim 40$ in the region we are interested in, we can set
approximately $u=\infty$, meaning no finite $\xi$ corection $G_L(u)=G_T(u)=1$,  
to obtain
\bq
F_L(x,Q^2)=  {{Q^2}\over{4\pi^2\alpha}} \frac{\sigma_{\gamma p}(W^2)} { \Bigl(\ln {\rho\over{\mu(W^2)}}\Bigr) }
I_L^{(1)}(\eta,\mu))
\label{a.15}
\eq
in (\ref{4.8}) of the main text.

\setcounter{equation}{0}
\setcounter{section}{2}
\section*{Appendix B. Evaluation of $C_2$ in $\Lambda_{sat}^{2}(W^2)$}
We note that in the CDP, for $x\ll 0.1$ and $Q^2$ sufficiently large \cite{{Ku-Schi}},
the (unmodified) evolution equation (\ref{4.3}) becomes 
\bq 
{\partial\over{\partial \ln W^2}}F_2\Bigl({{\xi_L}\over{\xi_2}}W^2\Bigr) 
= {1\over{2\rho_W +1}}F_2(W^2).
\label{b.1}
\eq
We remark that the hadronic ratio of longitudinal-to transverse $(q\bar q)_{L,T}~ p$ 
scattering, $\rho_W$, is at most weakly dependent on W, i.e. $\rho_W=\rho\cong {\rm const}
=4/3$.  Inserting 
\bq
F_2(W^2)\propto \Lambda^2_{sat}(W^2)=C_1\Bigl({{W^2}\over{1{\rm GeV}^2}}\Bigr)^{C_2},
\label{b.2}
\eq
into (\ref{b.2}), we obtain an important constraint,
\bq
C_2 (2 \rho + 1) \left( \frac{\xi_L}{\xi_2} \right)^{C_2} = 1,
\label{b.3}
\eq 
or conveniently to first order in $C_2$
\bq
C_2 \cong \frac{1}{2 \rho + 1} \frac{1}{\left( 1- \frac{1}{2 \rho + 1} \ln 
	\frac{\xi_2}{\xi_L} \right) }.
\label{b.4}
\eq 
With the preferred value of ${{\xi_2}\over{\xi_L}}=1.25$ and  $\rho={4\over 3}$,
one obtains the  prediction  of
\bq
C_2 =0.29.
\label{b.5}
\eq  
Evolution  accordingly implies the exponent of $\Lambda^2_{sat}(W^2)$
proportional to $\Bigl( \frac{W^2}{1{\rm GeV}^2} \Bigr)^{C_2}$   
consistent with its experimental determination .

%%%%%%%%%%%%%%%%%%%%%%%%%%%%%

\end{document}